\documentclass[twocolumn,prd, superscriptaddress,altaffilletter,nofootinbib, aps]{revtex4-1}
\usepackage{amsmath}
\usepackage{amssymb}
\usepackage{amsfonts}
\usepackage{comment}
\usepackage{graphicx}
\usepackage[colorlinks=true,citecolor=blue,urlcolor=blue]{hyperref}
\usepackage{acronym}
\usepackage[usenames,dvipsnames]{xcolor}
\usepackage{xspace}
\usepackage{tikz}
\usepackage{siunitx}
\usepackage{makecell}
\usepackage{multirow}
\usepackage{bm}
\usepackage{xcolor}
\usepackage{orcidlink}
\usepackage[caption=false]{subfig}
\usepackage{ulem}
\usepackage{booktabs}
\usepackage{physics}
\usepackage{tensor}
\usepackage{braket}

\usetikzlibrary{arrows,shapes,trees,decorations.pathreplacing}
\allowdisplaybreaks

\newcommand{\lvk}[0]{\ac{LVK}\xspace}
\newcommand{\gw}[0]{\ac{GW}\xspace}
\newcommand{\gws}[0]{\acp{GW}\xspace}
\newcommand{\bbh}[0]{\ac{BBH}\xspace}
\newcommand{\bbhs}[0]{\acp{BBH}\xspace}
\newcommand{\bns}[0]{\ac{BNS}\xspace}
\newcommand{\bnss}[0]{\acp{BNS}\xspace}

\newcommand{\cbc}[0]{\ac{CBC}\xspace}
\newcommand{\cbcs}[0]{\acp{CBC}\xspace}
\newcommand{\snr}[0]{\ac{SNR}\xspace}
\newcommand{\snrs}[0]{\acp{SNR}\xspace}
\newcommand{\SSM}{SSM\xspace}

\newcommand{\ICRR}{Institute for Cosmic Ray Research, The University of Tokyo, 5-1-5 Kashiwanoha, Kashiwa, Chiba 277-8582, Japan}

\begin{document}

\title{Efficient Reconstruction of Matched-Filter Signal-to-Noise Ratio Time Series from Nearby Templates for Compact Binary Coalescences Searches}

\author{Yasuhiro Murakami~\orcidlink{0009-0006-3400-057X}}
\affiliation{\ICRR}
\author{Tathagata Ghosh~\orcidlink{0000-0001-9848-9905}}
\affiliation{\ICRR}
\author{Soichiro Morisaki~\orcidlink{0000-0002-8445-6747}}
\affiliation{\ICRR}

\begin{abstract}

We present a method for efficiently searching long-duration gravitational wave signals from \cbcs. The approach exploits the smooth frequency-domain behavior of ratios between neighboring waveform templates. The matched-filter \snr time series of a data segment is first computed for a reference template, and the \snrs of nearby templates are then reconstructed by convolving this reference \snr time series with the ratio waveforms, defined as the frequency-domain ratios between the reference and neighboring templates. The computational speedup arises because the ratio waveforms can be safely truncated: they are significant only over a short interval approximately equal to the duration difference between the templates.
Storing these truncated ratio waveforms is practical and enables additional efficiency gains, in contrast to storing full templates, which is generally infeasible for long-duration, low-mass signals.
We demonstrate the efficacy of the method with mock non-spinning \cbc injections in the $1–3~M_\odot$ range. The reconstructed \snr time series agrees with that obtained from standard matched filtering to an accuracy of $\mathcal{O}(10^{-4})$, while the relative computational cost is reduced by $\gtrsim 25\%$. With a truncation threshold of $10^{-3}$ applied to the ratio waveform amplitudes, the storage requirement is reduced by a factor of $\sim 60$ relative to storing the full template bank.

\end{abstract}

\date{\today}

\maketitle

\acrodef{LVK}{the LIGO--Virgo--KAGRA collaboration}
\acrodef{GW}{gravitational wave}
\acrodef{BBH}{binary black hole}
\acrodef{BH}{black hole}
\acrodef{BNS}{binary neutron star}
\acrodef{NSBH}{neutron star black hole}
\acrodef{CBC}{compact binary coalescence}
\acrodef{SNR}{signal-to-noise ratio}

\section{Introduction}

The first detection of \gws from the \bbh merger GW150914~\cite{LIGOScientific:2016aoc} opened a new observational window onto the Universe, inaugurating the era of \gw astronomy. About two years later, \gws from a \bns merger, GW170817, were detected \cite{LIGOScientific:2017vwq}, accompanied by electromagnetic counterparts spanning radio to gamma rays \cite{LIGOScientific:2017ync}, establishing the first multi-messenger \gw event. Subsequent observations also identified \gws from neutron star black hole mergers \cite{LIGOScientific:2021qlt, LIGOScientific:2024elc}. To date, \lvk has reported hundreds of compact binary coalescence (CBC) events \citep{LIGOScientific:2018mvr,LIGOScientific:2020ibl,KAGRA:2021vkt,LIGOScientific:2025slb}, enabling us to learn the population properties of merging compact binaries \cite{LIGOScientific:2018jsj, KAGRA:2021duu, LIGOScientific:2025pvj}.

The primary method used in current \cbc searches is matched filtering~\cite{Sathyaprakash:1991mt, Dhurandhar:1992mw}. In this method, the detector data are cross-correlated with a bank of theoretically modeled waveforms, known as templates. The template bank must densely cover the parameter space of possible \cbcs, such that any astrophysical signal will have sufficiently high overlap with at least one template \cite{Owen:1995tm,Owen:1998dk}. In practice, the template bank is constructed to satisfy a predefined minimal match criterion (typically 0.97), which ensures that the loss in \snr due to mismatch between a true signal and the nearest template does not exceed a few percent. 

In this paper, we propose a technique to reduce the storage requirements and computational cost of matched filtering. Our method is particularly beneficial for low-mass \cbcs, such as \bnss and subsolar-mass (\SSM) \bbhs, whose long-duration signals lead to long template waveforms and expanded template banks. This issue becomes more critical for next-generation detectors such as \textit{Einstein Telescope}~\citep{Punturo:2010zz} and \textit{Cosmic Explorer}~\citep{Reitze2019Cosmic}, whose enhanced low-frequency sensitivity further increases signal durations and, consequently, the cost of matched filtering.

Our method builds upon ideas similar to Heterodyning \citep{Cornish:2010kf,Cornish:2021wxy} and Relative Binning \citep{Zackay:2018qdy,Krishna:2023bug}, which were developed for accelerating \cbc parameter estimation. These techniques exploit the fact that neighboring frequency-domain waveforms in parameter space differ mainly by smooth, slowly varying factors, exhibiting far fewer oscillations than the original waveforms. In these approaches, a fiducial waveform close to the true signal is selected, and the smooth ratio between each target waveform and the fiducial one is interpolated. As a result, the waveform need not be evaluated at every frequency during likelihood computation, greatly reducing the number of waveform evaluations required.

In this work, we extend the central ideas of the Heterodyning and Relative Binning approaches to perform matched filtering, which we refer to as \textit{ratio filter}.
We define a \textit{reference waveform} $\tilde{h}_{\mathrm{ref}}(f)$, serving as a fiducial waveform representative of a local region in parameter space.
For each region, we compute the matched-filter \snr time series using the reference waveform, denoted as $z_{\mathrm{ref}}(t)$.
The matched-filter \snr time series of a nearby waveform $\tilde{h}(f)$ can then be efficiently obtained from $z_{\mathrm{ref}}(t)$ by utilizing the waveform ratio $\tilde{r}(f) = \tilde{h}(f) / \tilde{h}_{\mathrm{ref}}(f)$.
Let $r(t)$ denote the inverse Fourier transform of $\tilde{r}(f)$.
It can be shown that the \snr time series of $\tilde{h}(f)$ is given by the convolution of $z_{\mathrm{ref}}(t)$ with $r(t)$.

Because $\tilde{r}(f)$ varies smoothly with frequency, its inverse Fourier transform $r(t)$ has a much shorter effective duration than the original waveforms, approximately equal to the difference between those of the two original waveforms.
This property allows matched filtering to be performed using these shorter ratio templates while retaining the full sensitivity of the original signals.
Since only a shorter waveform than original is sufficient for correlation, waveform generation becomes faster, and the required memory footprint is greatly reduced, enabling efficient searches for long-duration \gw signals.
This approach is also useful for two-stage hierarchical search strategies~\cite{Gadre:2018wsb,Soni:2021vls}, in which the parameter space is first explored with a coarse grid to identify candidate triggers, followed by refined analyses over a denser parameter grid in the second stage.

This paper is organized as follows. In Sec. \ref{sec:method}, we describe the conventional matched-filtering method and our ratio-filter technique. In Sec.~\ref{sec:simulation}, we outline the simulation setup used to validate our method.
The results are presented in Sec.~\ref{sec:results}, and we conclude in Sec.~\ref{sec:conclusions}. Appendix~\ref{sec:app1} derives the approximate expression for $r(t)$ under the stationary phase approximation, while Appendix~\ref{appendix:snr_loss} provides additional results related to the accuracy of the method.

\section{Methodology} \label{sec:method}

In this section, we describe the formalism of the ratio-filter technique developed in this work.
To establish notation and for completeness, we begin by defining the Fourier transform. Let $a(t)$ denote a real-valued time-domain signal (e.g., a GW strain), its Fourier transform is defined as
\begin{equation}
    \Tilde{a}(f) = \mathcal{F}[a(t)] = \int_{-\infty}^\infty a(t) \text{e}^{-2\pi ift} \text{d}t ,
\end{equation}
and its inverse transform is denoted by $a(t)=\mathcal{F}^{-1}[\Tilde{a}(f)]$.

\subsection{Matched filtering} \label{sec:matched_filter}

The detection of \cbc signals relies on the matched-filtering technique, which computes matched-filter \snr $\rho(t)$ by cross-correlating the interferometer output with a modeled waveform. This modeled waveform used for the cross-correlation is commonly referred to as a template.
The matched-filter \snr $\rho(t)$ is the modulus of the complex \snr $z(t)$~\cite{Allen:2005fk}, defined as:
\begin{equation}
\label{eq:matched_filter}
    z(t) = \frac{4}{N} \int_{0}^{\infty} \frac{\Tilde{x}(f)\Tilde{h}^*(f)}{S_n(f)}e^{2\pi ift}\text{d} f,
\end{equation}
where $\Tilde{x}(f)$ and $\Tilde{h}(f)$ 
denote the Fourier transforms of the detector strain and the template waveform, respectively, and $S_n(f)$ represents the one-sided noise power spectral density (PSD) of the detector noise.
$N$ is a normalization factor given by
\begin{equation}
    N = \sqrt{4\int_{0}^{\infty} \frac{|\Tilde{h}(f)|^2}{S_n(f)} \text{d} f.}
\end{equation}
In practical analyses, the integration range is limited by the detector’s sensitive frequency band, denoted by $[f_{\rm low}, f_{\rm high}]$. To reflect this, we assume
\begin{equation}
    S_n(f)=+\infty, \quad (f<f_{\text{low}} \quad\text{or}\quad f_{\text{high}}<f)
\end{equation}
so that the integration is performed in the meaningful range.

\subsection{Principle of ratio filter} \label{sec:principle}

In this section, we describe our ratio-filter technique, which enables interpolation of the matched-filter \snr time series for neighboring template parameters.
Let $\Tilde{h}_{\rm ref}(f)$ be the reference waveform, for which the complex matched-filter \snr time series $z_{\rm ref}(t)$ has been computed, and $\Tilde{h}(f)$ denote a nearby waveform (hereafter referred to as the target waveform).
These waveforms are assumed to have comparable signal durations. By taking the ratio, we define
\begin{equation}\label{eq:ratio_waveform}
    \Tilde{r}(f) \equiv \frac{\Tilde{h}(f)}{\Tilde{h}_{\rm ref}(f)}, \quad r(t) \equiv \mathcal{F}^{-1}[\Tilde{r}(f)].
\end{equation}
The complex matched-filter \snr time series associated with $\Tilde{h}_{\rm ref}(f)$ is given by
\begin{align}
    z_{\rm ref}(t) &= \frac{4}{N_{\rm ref}} \int_0^\infty \frac{\Tilde{x}(f)\Tilde{h}^*_{\rm ref}(f)}{S_n(f)}e^{2\pi ift}\text{d}f \notag \\
    &= \int_0^\infty \Tilde{z}_{\rm ref}(f) e^{2\pi ift}\text{d}f, \label{eq:snr_ref}
\end{align}
where $\Tilde{z}_{\rm ref}(f)$ represents the frequency-domain components of $z_{\rm ref}(t)$ and $N_{\rm ref}$ is a normalization factor associated with reference waveform $\Tilde{h}_{\rm ref}(f)$. 
Using $\Tilde{z}_{\rm ref}(f)$ and $\tilde{r}(f)$, we can express the complex matched-filter \snr timeseries associated with the target waveform as
\begin{align}
    z(t) &= \frac{4}{N} \int_0^\infty \frac{\Tilde{x}(f)\Tilde{h}^*(f)}{S_n(f)}e^{2\pi ift}\text{d}f \notag \\
    &= \frac{N_{\rm ref}}{N} \int_0^\infty \Tilde{z}_{\rm ref}(f) \Tilde{r}^*(f) e^{2\pi ift}\text{d}f. \label{eq:related_binning}
\end{align}
Since the product in Fourier space corresponds to convolution, this can also be written as
\begin{equation} \label{eq:convolution}
    z(t) = \frac{N_{\rm ref}}{N} \int^\infty_{-\infty} z_{\rm ref}(t+s) r(s) \text{d}s.
\end{equation}
Thus, the target \snr timeseries $z(t)$ for a nearby template can be obtained by convolving the precomputed \snr time series $z_{\rm ref}(t)$ with $r(t)$.

\begin{figure*}
    \centering
    \includegraphics[width=1\hsize]{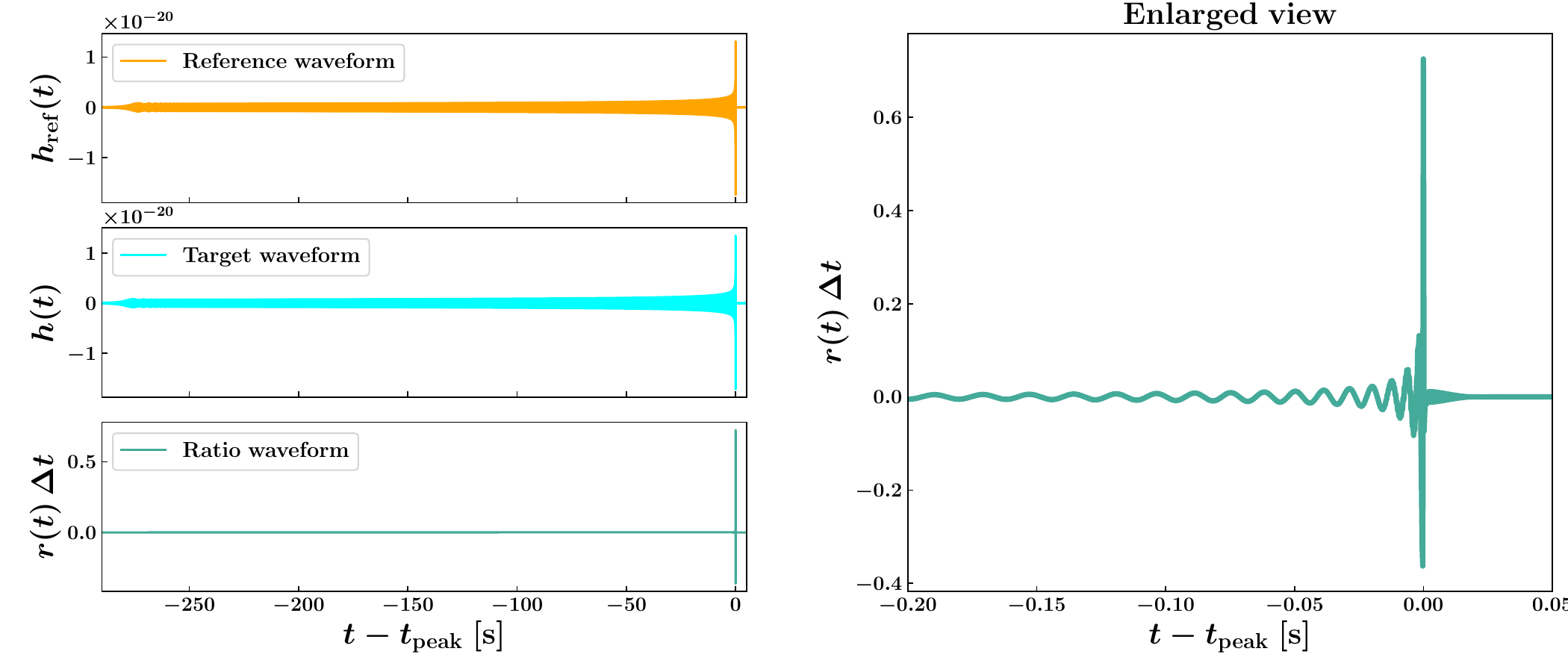}
    \caption{\textit{Left Panel:} Time-domain plots of the reference (top) and target (middle) waveforms generated using the \textsc{TaylorF2} model, and their ratio (bottom), all shown as a function of time relative to the waveform peak $t_\mathrm{peak}$. The reference waveform has a chirp mass $\sim0.875~M_\odot$ with a mass ratio of unity, while the target waveform has a chirp mass $\sim0.870~M_\odot$ with the same mass ratio.}
    \label{fig:waveforms}
\end{figure*}

The key advantage of this method is that the \textit{ratio waveform}, $r(t)$, is only significant within a narrow time window, much shorter than the original waveforms from which it is constructed. In fact, the effective duration is approximately equal to the difference between the durations of the two waveforms being compared, as illustrated in Fig.~\ref{fig:waveforms}. 
In the left panel, the top and middle curves show the reference and target waveforms, both generated using the \textsc{TaylorF2} waveform model~\cite{Buonanno:2009zt} and each with a duration of $\sim 300$ seconds. The bottom panel displays the corresponding ratio waveform $r(t)$, representing the ratio between the reference and target waveforms. 
Since the ratio waveforms are significant over a small time, they can be pre-generated and stored for later analysis.
Consequently, using the truncated ratio waveform in matched filtering substantially reduces computational cost by decreasing both the waveform length and the time required for waveform generation.
The formalism of truncating the ratio waveform is described in Sec.~\ref{sec:truncating_waveform}.

\subsection{Window functions} \label{sec:window_function}

For the ratio-filter method to work efficiently, $r(t)$ must be negligibly small outside a narrow time window. On the other hand, if $\tilde{r}(f)$ has abrupt cutoffs at $f_{\rm low}$ or $f_{\rm high}$, its inverse Fourier transform develops long tails that extend beyond the interval $\min \left[\delta t(f_{\rm low}), \delta t(f_{\rm high})\right] < t < \max \left[\delta t(f_{\rm low}), \delta t(f_{\rm high})\right]$ \cite{Morisaki:2021ngj}. To mitigate this effect, we apply the smooth window function as described below:
\begin{widetext}
\begin{equation}
\label{eq:window}
    W(f)= \begin{cases}\frac{1}{2}\left(1+\cos \left(\pi \frac{f-f_\text{low}}{\Delta f_\text{low}}\right)\right), & \left(f_\text{low}-\Delta f_\text{low}<f<f_\text{low}\right), \\ 
    1, & \left(f_\text{low} \leq f \leq f_\text{high}\right), \\ 
    \frac{1}{2}\left(1-\cos \left(\pi \frac{f-(f_\text{high} + \Delta f_{\text{high}})}{\Delta f_\text{high}}\right)\right), & \left(f_\text{high}<f<f_\text{high}+\Delta f_\text{high}\right), \\ 
    0, & \text { (otherwise). }\end{cases}
\end{equation}
\end{widetext}

\begin{figure}[h]
    \centering
    \includegraphics[width=1\hsize]{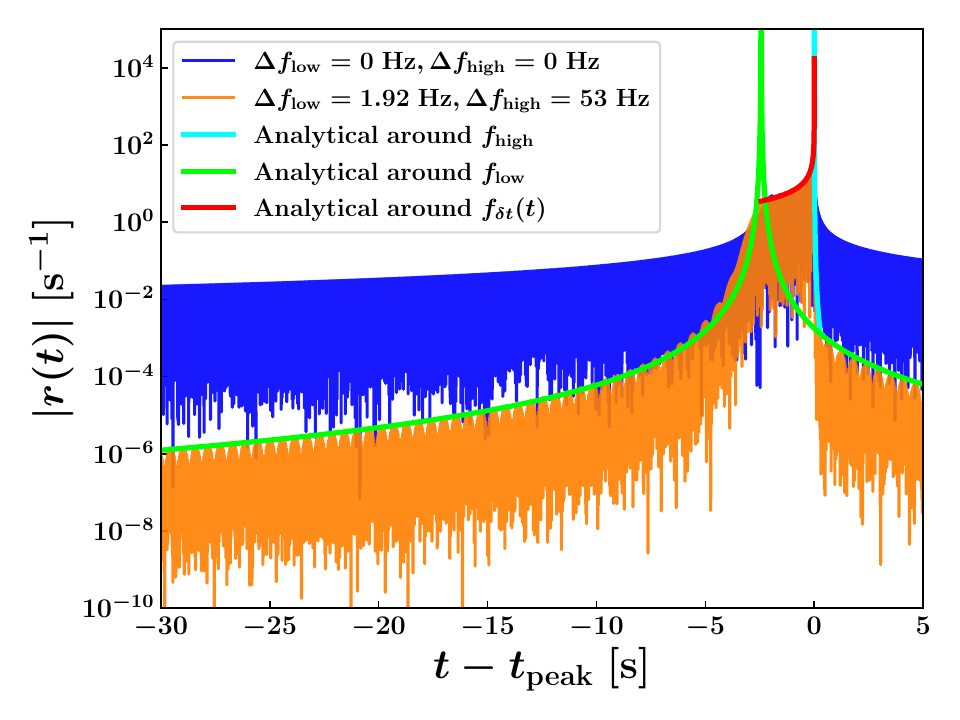}
    \caption{Absolute value of the ratio waveform $r(t)$ as a function of time relative to the waveform peak, $t-t_{\rm peak}$. The blue and orange curves corresponding to the ratio waveforms with the window function Eq.\eqref{eq:window} for $\Delta f_{\rm low}=0~{\rm Hz}, \Delta f_{\rm high}=0~{\rm Hz}$, and $\Delta f_{\rm low}=1.92~{\rm Hz}, \Delta f_{\rm high}=53~{\rm Hz}$, respectively. The ratio waveform is computed using Eq.\eqref{eq:ratio_waveform}, with the same parameters for the reference and target waveform as in Fig.~\ref{fig:waveforms}. The green, cyan, and red curves represent the analytic expressions for the upper bounds evaluated at $f_\text{high}, f_\text{low}$, and $f_{\delta \tau}(t)$, as derived in Appendix~\ref{sec:app1}.}
    \label{fig:ratio_waveform}
\end{figure}

A simple truncation of the waveform in the frequency domain without windowing can leave residual tails in the time domain (see Fig.~\ref{fig:ratio_waveform}), which degrades the accuracy of the analysis.
Here, $\Delta f_\text{low}$ is chosen as
\begin{equation}
    \Delta f_\text{low} = \frac{1}{\sqrt{|\delta\tau'(f_{\text{low}})|}}.
\end{equation}
Here, $\delta\tau(f) = \tau(f) - \tau_{\text{ref}}(f)$ and 
$'$ represents derivative with respect to frequency where $\tau(f)$ denotes the duration until merger from frequency $f$, and $\tau_{\text{ref}}(f)$ is the corresponding value for the reference waveform.
However, if $|\delta\tau'(f_{\text{low}})|$ is too small, $\Delta f_\mathrm{low}$ becomes very large, so in this paper, the maximum value of $\Delta f_\mathrm{low}$ is set to 5 Hz. For $\Delta f_\text{high}$, following Ref.~\cite{Morisaki:2021ngj}, we take $\Delta f_\text{high}=53$ Hz.
In Ref.~\cite{Morisaki:2021ngj}, this value was chosen to ensure that the time-domain waveform quickly attenuates after the coalescence time. It effectively vanishes by the end of the data segment, which was set to $\sim 2\,\mathrm{s}$ after the coalescence time. In general, $\Delta f_\text{high}$ depends on the offset between the coalescence time and the end of the analyzed data, and may therefore differ in our case. We defer the optimization of $\Delta f_\text{high}$ to future work.

\subsection{Truncating the ratio waveform}\label{sec:truncating_waveform}

We have discussed in Sec.~\ref{sec:principle} that the ratio waveform $r(t)$ is only significant over a narrow time interval. Therefore, in evaluating Eq.\eqref{eq:convolution}, it is sufficient to restrict the computation to the effective time window. Since the behavior of $r(t)$ can be characterized analytically, the appropriate truncation range can be determined from the analytical expression (see Appendix~\ref{sec:app1}).

Although the reference and target waveforms slightly differ in their durations, they must be generated with the same frequency resolution $\Delta f$, since the inverse of $\Delta f$ sets the corresponding time-domain length.
Using a common frequency resolution ensures that their frequency samples are aligned, making it possible to compute the ratio waveform, as in Eq.~\eqref{eq:ratio_waveform}.
Moreover, since the inverse of $\Delta f$ determines the time-domain duration, it should be chosen to be at least longer than $\delta \tau(f_\mathrm{low})$.

To truncate the ratio waveform, we introduce a threshold amplitude and retain only the region where $|r(t)|$ exceeds this threshold using the analytical formula derived in Appendix \ref{sec:app1}.
In doing so, we exclude the low-amplitude regions of $r(t)$ that contribute negligibly to the matched filtering.

\begin{figure}
    \centering
    \includegraphics[width=1\hsize]{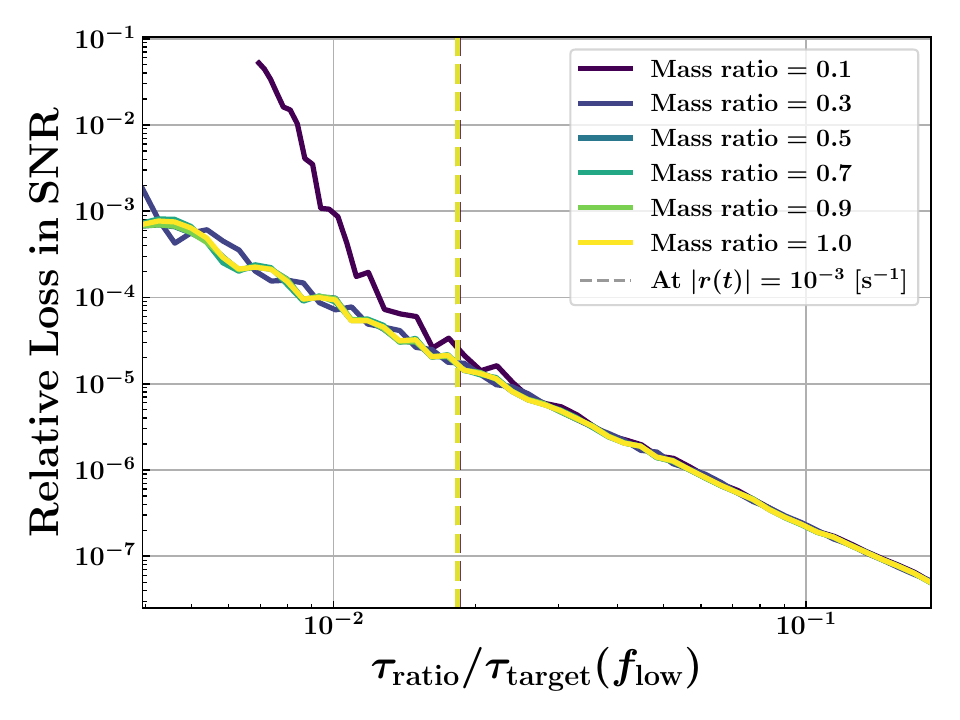}
    \caption{Relative \snr loss as a function of the retained side duration of the ratio waveform $r(t)$. Each curve corresponds to a different mass ratio, while keeping the chirp mass fixed at $1.219~M_\odot$. The horizontal axis represents the duration of the ratio waveforms $\tau_\mathrm{ratio}$, normalized by the duration of the target waveform $\tau_\mathrm{target}$, i.e., the duration of the injected signal. The ratio waveform length is varied by changing the threshold amplitude applied to $\abs{r(t)}$ using analytical solutions (Appendix~\ref{sec:app1}). The vertical dashed lines indicate the value of the horizontal axis when the threshold amplitude is set to $10^{-3}~[\mathrm{s^{-1}}]$.}
    \label{fig:relative_error}
\end{figure}

Fig.~\ref{fig:relative_error} shows the relative \snr loss obtained from an injection analysis designed to evaluate the accuracy of the proposed method. In this test, we inject a waveform with a fixed chirp mass and various mass ratios of $q=0.1-1.0$ into Gaussian noise, which is generated using the \textsc{aLIGOZeroDetHighPower} power spectral density (PSD) model \citep{ligo-sensitivity} representing the ideal LIGO design sensitivity. The injected data are analyzed using both the conventional matched filtering method and our ratio-filter method, and the relative \snr error between the two results is calculated. The analysis uses templates with the same physical parameters as those used for the injection, and determines the duration of $r(t)$, $\tau_\mathrm{ratio}$, by changing the threshold amplitude. In the figure, the mass ratio $q$ shown for each curve corresponds to the injection waveforms, while the reference waveform used in the ratio-filter method is fixed to $q=0.6$. The relative error decreases as the retained portion of the waveform becomes longer, since more information is preserved. However, a longer retained waveform also increases the computational cost, indicating a clear trade-off between accuracy and efficiency. In this paper, we adopt a threshold amplitude of $10^{-3}~[\mathrm{s^{-1}}]$ as a practical balance between these two factors. This corresponds to the vertical dashed lines shown in Fig.~\ref{fig:relative_error}.
The behavior of relative \snr loss for different choices of chirp masses is discussed in Appendix~\ref{appendix:snr_loss}.

\section{Simulation} \label{sec:simulation}

To assess the efficacy of our method described in Sec.~\ref{sec:method}, we apply both standard matched filtering and ratio filter to simulated data. We focus on binary neutron star mergers with component masses in the range $1$–$3~M_{\odot}$. A non-spinning template bank is constructed and further organized into reference templates and their corresponding dependent templates, as detailed in Sec.~\ref{sec:template_bank}. The structured template bank is essential for ratio filtering, whereas standard matched filtering can be performed directly without such organization. Details of the simulated data used for this analysis are presented in Sec.~\ref{sec:mock_data}.

\subsection{Template Bank Construction}\label{sec:template_bank}

We first construct the template bank and then organize it into reference templates and their corresponding dependent templates. 
We first perform matched filtering with the reference waveforms (Eq.\eqref{eq:matched_filter}). Once the SNR time series for all reference waveforms is obtained, we compute the SNR time series of the dependent templates associated with each reference template using Eq.\eqref{eq:convolution}.

We construct a non-spinning template bank in the mass range $1$–$3M_{\odot}$ using a stochastic placement algorithm, ensuring that any two neighboring templates have a minimal match of $0.97$. The match is calculated in a transformed parameter space, as proposed in Ref.\cite{Brown:2012qf}, rather than in the source parameters (e.g., $m_1$, $m_2$, etc.), such that the metric in the new parameter space is globally flat. 
With this choice, the template bank used in this study contains approximately $5.1 \times 10^{4}$ templates.

When structuring the template bank, we must ensure that within each set of templates -- consisting of one reference waveform and its corresponding dependent waveforms -- the variation in signal duration is not too large. We fix this threshold at a fiducial value of $10$ seconds. Since the chirp mass primarily determines the waveform duration, we employ a bisection procedure: starting from the complete chirp-mass ordered templates, we iteratively split the set into two halves until, for every subset, the difference between the longest and shortest waveform durations is equal to or less than $10$ seconds.
Since the effective duration of the ratio waveform is approximately similar to that of the two waveforms used to compute it, we adopt this value to keep the waveform limited in duration and storage requirements reasonable.
We do not focus on optimizing this choice, which may further improve the efficiency of the method. We leave such exploration for future work.

The structured template bank exhibits significant variation in the number of templates across different sets. To reduce this variation and obtain subsets with a roughly similar number of templates, we further subdivide each set by performing a bisection in symmetric mass ratio, continuing the process until each subset contains no more than $500$ templates. However, exact uniformity can not always be achieved, since we must also ensure that the duration difference between the shortest and longest waveforms remains within $10$ seconds. The fiducial values, corresponding to a maximum of $500$ templates per subset and a $10$ seconds limit on the duration difference between the shortest and longest waveform, can be refined in future work to achieve optimal performance.

Finally, we identify a reference waveform for each set of templates. For a given set, the reference waveform is chosen as the one that minimizes the sum of the absolute differences in duration when compared with all other templates in that set.
We then compute the ratio waveforms used for constructing the SNR time series, following the procedure introduced in Sec.~\ref{sec:method}.

\begin{figure*}
    \centering
    \includegraphics[scale=0.58]{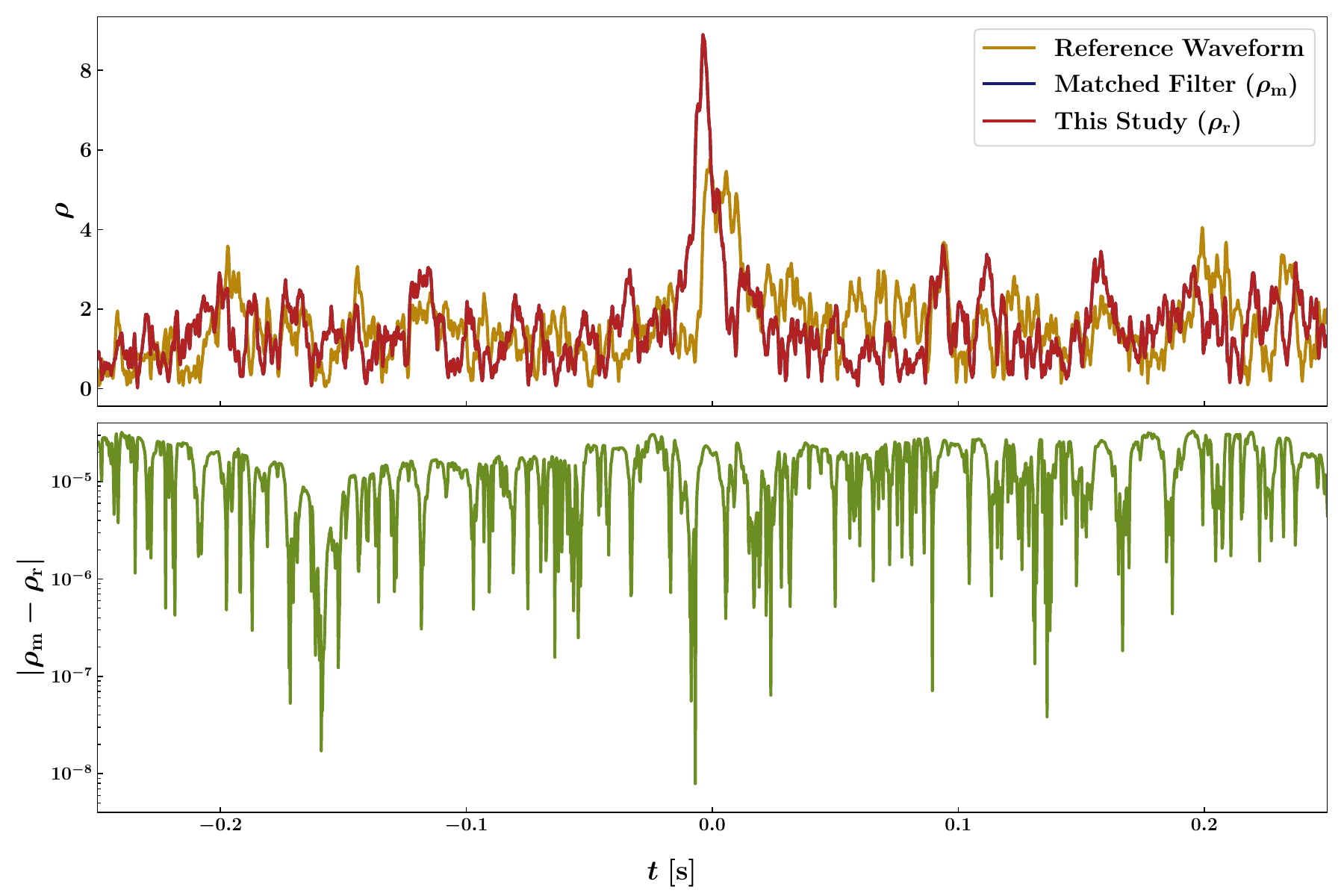}
    \caption{
    \textit{Top Panel:} SNR time series of the reference waveform associated with the group containing the loudest template among its dependent templates.
    Both $\rho_m(t)$ and $\rho_r(t)$ correspond to the same template which yields the highest recovered \snr in the template bank.
    \textit{Bottom Panel:} Comparison of the SNR time series from standard and ratio filter.}
    \label{fig:snr_ts}
\end{figure*}

\subsection{Mock Data Preparation} \label{sec:mock_data}

We use the \textsc{aLIGOZeroDetHighPower} PSD~\cite{ligo-sensitivity} to generate realistic noise. The simulated noise has a duration of $512$ seconds, sampled at $4096$~Hz, and spans the frequency range of $20$–$2048$~Hz. This duration is sufficient to accommodate the injected signals within the mass range considered in this study. Simulated non-spinning CBC signals are then injected into the colored Gaussian noise for the LIGO Hanford (H1) detector. The same waveform model, \textsc{TaylorF2}, used for constructing the template bank, is used to generate the CBC signals, and their details are provided in Sec.~\ref{sec:results}. The sky positions of the sources are randomly distributed over the entire sky. 
The luminosity distance for each injection is chosen such that the resulting signal in zero noise has an optimal matched-filter SNR of $8$ in the H1 detector if we consider the same signal as a template.
However, when the same signal is injected into colored Gaussian noise, the recovered SNR at the injection time fluctuates around this value due to the additive noise. The simulated signals neglect tidal deformability, as our focus is on testing the filtering performance rather than waveform systematics.

\section{Results} \label{sec:results}

Once the mock data are prepared, we perform both the standard matched filtering procedure and our ratio-filter method on the same datasets. Before performing the ratio filter, the ratio waveforms are computed following the procedure described in Sec.~\ref{sec:method}, and stored for subsequent use. 
The ratio-filtering proceeds by first calculating the reference SNR time series using the reference templates, as given in Eq.~\eqref{eq:snr_ref}. 
These reference SNR time series are then used to compute the SNR time series of the neighboring templates by convolving them with the corresponding ratio waveforms, as given in Eq.~\eqref{eq:convolution}. The convolution is carried out using the \textsc{oaconvolve} routine implemented in the \textsc{scipy} \cite{2020SciPy-NMeth}.
It is important to note that we have zero-padded the signal above $2048$~Hz by an additional $53$~Hz, as the ratio filters are defined over the frequency range $15$–$2101$~Hz (see Sec.~\ref{sec:window_function}). For consistency, the same zero-padded signal is used when computing the standard matched-filter SNR. Note that the matched-filter SNR is evaluated over $20$–$2048$~Hz.
 
\begin{figure}
    \centering
    \includegraphics[scale=0.56]{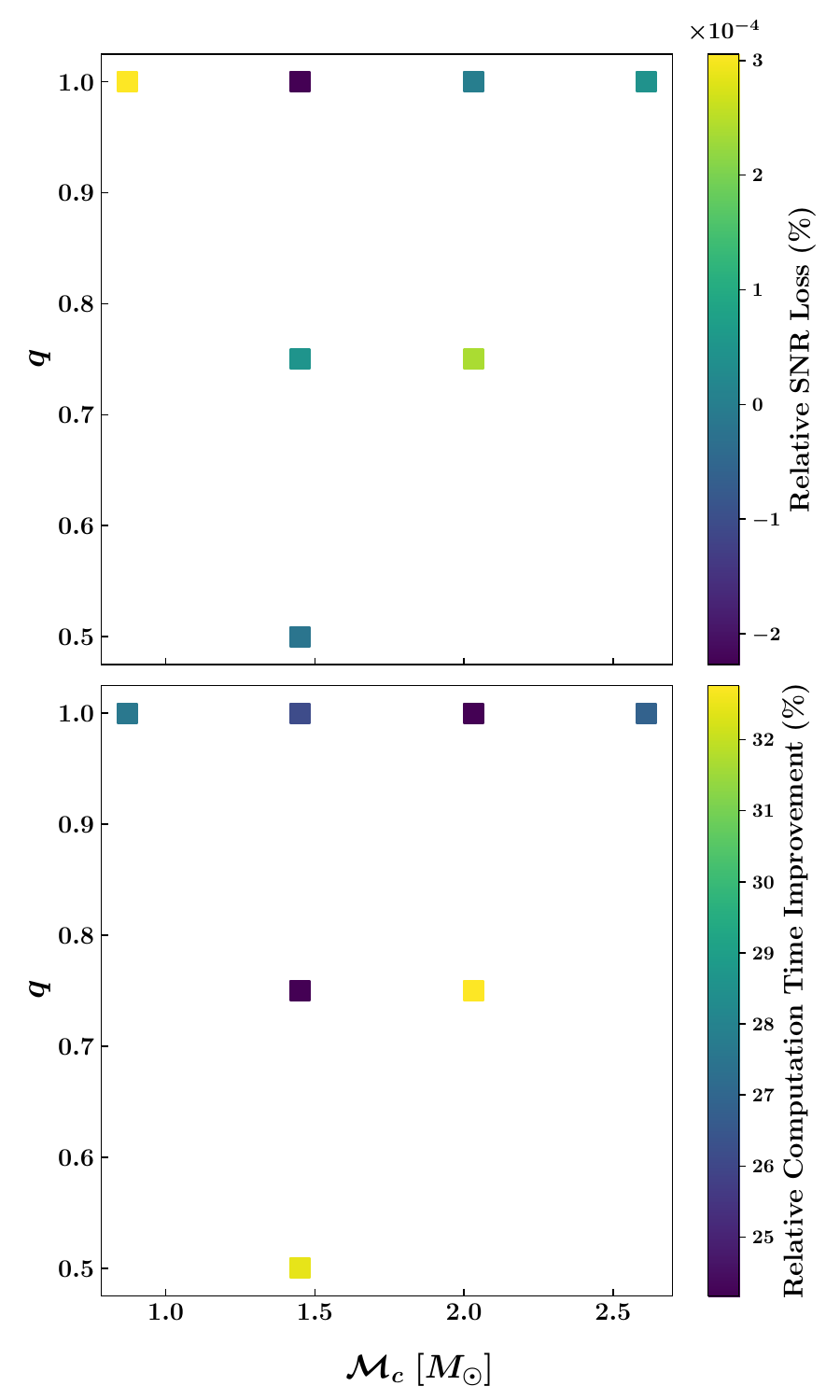}
    \caption{Comparison of the relative SNR loss (top) and improvement in computation time (bottom) between standard and ratio filter.}
    \label{fig:mf_performance}
\end{figure}

As a representative case, Fig.~\ref{fig:snr_ts} presents the SNR time series obtained from our analysis for a $(1.67+1.67)~M_{\odot}$ BNS merger~\footnote{All masses are quoted in the detector frame.}. The top panel shows the SNR time series of the reference waveform corresponding to the group that contains the loudest template among its dependent templates. 
The matched-filter SNR time series corresponding to the reference waveform shows a peak value of $5.769$ for a CBC signal whose zero-noise optimal SNR is set to $8$. 
However, when the signal is injected into colored Gaussian noise, the recovered SNR is slightly higher due to the specific noise realization. This is evident in the top panel of Fig.~\ref{fig:snr_ts}, which compares the standard matched-filter SNR time series with that obtained using our method. The recovered matched-filter SNRs agree well, with both yielding a value of $8.903$, and the absolute difference between the two estimates is shown in the bottom panel.

We consider a few representative mass parameters (one such example is discussed above) to investigate the performance of the ratio filter compared to standard matched filtering in terms of SNR loss and computational cost. Specifically, we use four equispaced points in the chirp-mass plane between $0.8705~M_{\odot}$ and $2.612~M_{\odot}$, corresponding to $(1+1)~M_{\odot}$ and $(3+3)~M_{\odot}$ BNS mergers, assuming equal-mass binaries. The mass ratios are chosen as $0.5$, $0.75$, and $1$, subject to the constraint that the component masses remain within $1$–$3~M_{\odot}$, consistent with the template bank used in this work. 
The top panel of Fig.~\ref{fig:mf_performance} shows the relative SNR loss incurred by the ratio filter compared to the standard matched filtering, with deviations at the level of $\mathcal{O}(10^{-4})\%$. 
The bottom panel of Fig.~\ref{fig:mf_performance} shows the relative improvement in computation time achieved by the ratio filter over the standard approach, with a gain of  $\sim 25\%$.
For standard matched filtering, this includes waveform generation and the cross-correlation. For the ratio filter, the timing includes the matched filtering of the reference templates with the data segment and the convolution of the ratio waveforms with the reference SNR time series.

It is important to note that, unlike in the standard matched filtering, where the waveform is generated on the fly during the cross-correlation of templates with data, the ratio filter stores the ratio waveforms in advance. While storing the full waveforms in the standard approach could, in principle, reduce computation time significantly, it is generally challenging due to the substantial storage requirements. In contrast, the ratio waveforms require $\sim 60$ times less storage owing to their significantly shorter durations.~\footnote{We quantify this improvement by comparing the durations of the ratio and original waveforms in the template bank.}
Other techniques, such as singular value decomposition (SVD) \cite{Cannon:2010qh}, can also reduce the total template size. However, our method requires no offline SVD computation, and constructing ratio filters is straightforward.

\section{Conclusions} \label{sec:conclusions}

We have presented a method for efficiently searching for long-duration \gw signals from \cbc. Our approach builds on the ideas of heterodyning \citep{Cornish:2010kf, Cornish:2021wxy} and relative binning \citep{Zackay:2018qdy}, by exploiting the smooth frequency-domain behavior of ratios between neighboring waveform templates.
In practice, we compute the matched-filter \snr time series for each reference template in the parameter-space groups using the standard matched-filtering procedure. 
The \snr time series for the remaining (target) templates in each group is then reconstructed by convolving the reference \snr time series with the corresponding ratio waveforms, where each ratio waveform is defined as the frequency-domain ratio between the target and reference templates [see Eq.~\eqref{eq:ratio_waveform}].
This approach preserves the accuracy of the matched-filter \snr estimate, with differences relative to the standard procedure at the level of $\mathcal{O}(10^{-4})$.
One can improve the efficiency of the standard matched-filter procedure by storing waveform templates rather than generating them on the fly. However, for low-mass \cbcs, the templates are long in duration, and storing a full template bank requires substantial storage. 
In contrast, our method stores only the significant segment of the ratio waveforms, which spans a short interval approximately equal to the duration difference between the target and reference templates.
Consequently, the storage requirement is much smaller. 
For the mass range of $1$–$3~M_\odot$ considered in this study and using an amplitude threshold of $10^{-3}$, the storage required for the ratio waveforms is reduced by a factor of $\sim 60$ compared to storing the full templates.

In this study, we have demonstrated the method as a proof of concept using non-spinning injections. The framework, however, can be straightforwardly extended to include spin. Incorporating spin will increase the size of the template bank and, therefore, requires careful restructuring to optimally identify reference templates and their corresponding dependent templates. While we adopted one specific strategy for this restructuring (see Sec.~\ref{sec:template_bank}), alternative strategies may provide additional efficiency gains. We leave the exploration of these possibilities to future work.

This approach offers significant advantages for low-mass binary searches in future detectors such as the Einstein Telescope and Cosmic Explorer, where extending searches to lower frequencies can increase signal durations to several thousand seconds.
Similarly, searches for \SSM binaries with current-generation detectors face comparable challenges, as their long inspiral signals require much larger template banks. To mitigate this computational burden, the \lvk Collaboration has adopted relatively high low-frequency cutoffs (e.g., $45$ Hz) in the $[0.2,1]~M_{\odot}$ mass range, albeit at the cost of sensitivity \citep{LIGOScientific:2018glc, LIGOScientific:2019kan, LIGOScientific:2021job}. In this context, our method offers a promising avenue to pursue \SSM searches that could incorporate lower-frequency data. 

An alternative method based on Multi-Band Template Analysis (MBTA) \citep{Vinciguerra:2017ngf, Adams:2015ulm, Allene:2025saz} is employed in the \lvk low-latency \bns search to reduce the computational cost of matched-filtering. Multi-band techniques are also used in other low-latency pipelines \citep{Cannon:2011vi}. MBTA achieves this by splitting the frequency band into multiple sub-bands with different sampling rates and thereby reducing the computational cost per template while preserving the recovered \snr. In contrast, the ratio-filter method computes the matched-filter \snr time series only for a sparse set of reference templates, and reconstructs the \snrs of neighboring templates from the corresponding short ratio waveforms. The two approaches address different sources of computational cost and are therefore complementary. In principle, MBTA can be applied within each reference template, while the ratio-filter method reduces redundancy across templates. This combination thus has the potential to yield further acceleration for low-mass \cbc searches.

\acknowledgments
 
T.G. acknowledges using the LDG clusters at CIT and IUCAA (Sarathi) for the computation involved in this work. This work was supported by Japan Society for the Promotion of Science (JSPS) Grants-in-Aid for Transformative Research Areas (A) No.~23H04891 and No.~23H04893 (T.G., S.M.).

\section*{Data Availability}
The data are not publicly available. The data are available from the authors upon reasonable request.

%--------------Appendix----------------

\appendix

\section{Approximate expression for $r(t)$ under the stationary phase approximation}\label{sec:app1}

In this section, we derive an approximate expression for $r(t)$ using the stationary phase approximation (e.g., Ref.~\cite{Creighton:2011zz}).  
Under the stationary phase approximation, the frequency-domain gravitational waveform for a positive frequency $f$ can be written as
\begin{equation}
    \tilde{h}(f) = B(f)\, e^{-i\Psi(f)},
\end{equation}
where the derivatives of the phase are related to the time–frequency mapping $t(f)$ as
\begin{equation}
    \Psi'(f) = 2\pi t(f), \qquad \Psi''(f) = 2\pi t'(f).
\end{equation}
Accordingly, $\tilde{r}(f)$ for $f>0$ is approximated as
\begin{equation}
    \tilde{r}(f) = R(f) e^{-i \delta\Psi(f)},
\end{equation}
where
\begin{equation}
    R(f) \equiv \frac{B(f)}{B_{\rm ref}(f)}, \qquad \delta\Psi(f) \equiv \Psi(f) - \Psi_{\rm ref}(f),
\end{equation}
and $B_{\rm ref}(f)$ and $\Psi_{\rm ref}(f)$ represent the amplitude and phase of the reference waveform, respectively.

Hence, an approximate expression for $r(t)$ can be written as
\begin{equation}
    r(t) = 2\,\mathrm{Re} \left[ \int_{f_{\rm low}-\Delta f_{\rm low}}^{f_{\rm high}+\Delta f_{\rm high}} 
    \mathrm{d}f\, W(f)\, R(f) \, e^{2\pi i f t - i \delta\Psi(f)} \right].
\end{equation}
We evaluate this integral approximately following the approach described in Appendix~A of Ref.~\cite{Morisaki:2021ngj}.
Define $\delta t(f) = t(f) - t_{\rm ref}(f)$ for $f > 0$ and let $f_{\delta t}(t)$ denote its inverse, satisfying $\delta t(f_{\delta t}(t)) = t$.
Here, we assume that $\delta t(f)$ is a monotonic function.
The evaluation of the integral then depends on whether $f_{\delta t}(t)$ lies below $f_{\rm low}$, within the range $(f_{\rm low}, f_{\rm high})$, or above $f_{\rm high}$.

If $f_{\delta t}(t) \leq f_{\rm low}$, the dominant contribution to the integral arises from around $f=f_{\rm low}$.
Hence, we approximately fix the amplitude and expand the phase around $f=f_{\rm low}$ as follows,
\begin{align}
    &R(f) \simeq R(f_{\rm low}), \\
    &\begin{aligned}
        \delta \Psi(f) \simeq &\delta \Psi(f_{\rm low}) + 2\pi\delta t(f_{\rm low}) (f - f_{\rm low}) \\
        &+ \pi \delta t'(f_{\rm low}) (f - f_{\rm low})^2,
    \end{aligned}
\end{align}
extend the integral range to $[f_{\rm low}-\Delta f_{\rm low}, \infty]$, and assume the window function $W(f)$ is unity even above $f_{\rm high}$.
Following the same procedure to derive Eq. (15) of Ref.~\cite{Morisaki:2021ngj}, we obtain the following approximate expression, valid when $|t - \delta t (f_{\rm low})|$ is sufficiently large:
\begin{equation}
    \begin{aligned} \label{eq:flow_rezult}
        &r(t) \simeq \mathrm{Re} \Bigg[ \\
        &~~\frac{i}{8\pi}\frac{R(f_{\rm low})}{(\Delta f_{\rm low})^2(\delta t(f_{\rm low}) - t)^3} e^{2 \pi i f_{\rm low} t - i\delta\Psi(f_{\rm low})} \times \\
        &~~ \left(1 + e^{2\pi i \Delta f_{\rm low}(\delta t(f_{\rm low}) - t) - i\pi \left(\Delta f_{\rm low}\right)^2 \delta t'(f_{\rm low})}\right)\Bigg].
    \end{aligned}
\end{equation}
Hence, the amplitude is attenuated in proportion to $\left|\delta t(f_{\rm low}) - t\right|^{-3}$,
\begin{equation}
\label{eq:flow_result_amp}
    |r(t)| \leq \frac{1}{4\pi}\frac{\left|R(f_{\rm low})\right|}{(\Delta f_{\rm low})^2 \left|\delta t(f_{\rm low}) - t\right|^3}.
\end{equation}

Similarly, if $f_{\delta t}(t) \geq f_{\rm high}$, the dominant contribution to the integral arises from around $f=f_{\rm high}$.
Hence, we fix $R(f)$ to $R(f_{\rm high})$, expand $\Psi(f)$ around $f=f_{\rm high}$, extend the integral range to $[-\infty, f_{\rm high} + \Delta f_{\rm high}]$, and assume the window function $W(f)$ is unity even below $f_{\rm low}$.
Following the same derivation as above, we find that the amplitude decays in proportion to $\left|t - \delta t(f_{\rm high})\right|^{-3}$:
\begin{equation}
\label{eq:fhigh_result_amp}
    \left|r(t)\right| \leq \frac{1}{4\pi}\frac{\left|R(f_{\rm high})\right|}{(\Delta f_{\rm high})^2 \left|t - \delta t(f_{\rm high})\right|^3}.
\end{equation}

If $f_{\rm low} < f_{\delta t}(t) < f_{\rm high}$, the dominant contribution to the integral arises from around $f=f_{\delta t}(t)$.
Hence, we fix $R(f)$ to $R\left(f_{\delta t}(t)\right)$, expand $\Psi(f)$ around $f=f_{\delta t}(t)$, extend the integral range to $[-\infty, \infty]$, and assume the window function $W(f)$ is unity. 
The resulting approximate expression is
\begin{equation}
\begin{aligned}
\label{eq:delta_t_result}
    r(t) = &\Re \bigg[ \sqrt{2} \frac{(1 - i \mathrm{sgn}(\delta t'(f_{\delta t}(t))) R(f_{\delta t}(t))}{\sqrt{\abs{\delta t'(f_{\delta t}(t))}}} \\
    &\times e^{2 \pi i f_{\delta t}(t) t - i\delta\Psi\left(f_{\delta t}(t)\right)}\bigg].
\end{aligned}
\end{equation}

\section{Behavior of relative \snr loss for different chirp masses} \label{appendix:snr_loss}

\begin{figure*}
    \centering
    \includegraphics[width=1\hsize]{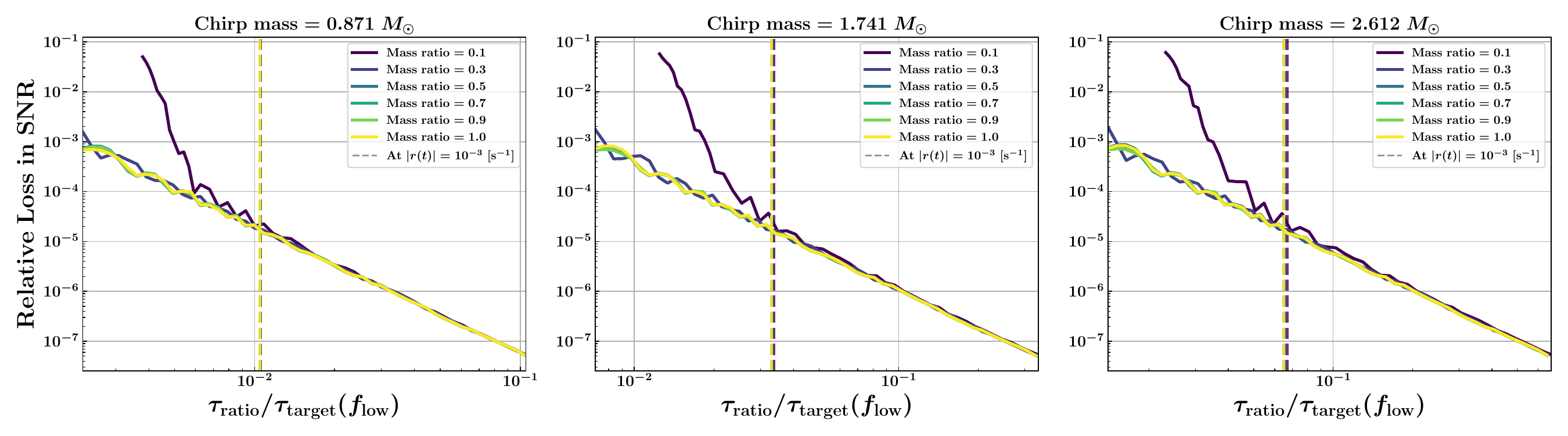}
    \caption{Same analysis as Fig.~\ref{fig:relative_error}, but for different chirp masses, $\mathcal{M}=0.871\,M_\odot$(left), $1.741\,M_\odot$(middle), and $2.612\,M_\odot$(right). The curves shift in the horizontal axis, but remain almost the same shape, and the \snr loss at the amplitude-based truncation threshold is consistent across all cases.}
    \label{fig:different_chirp_mass_error}
\end{figure*}

We examine how the \snr loss of the ratio waveforms depends on chirp mass as a function of 
$\tau_\mathrm{ratio}/\tau_\mathrm{target}$, as previously defined.
For a fixed chirp mass, we find that the overall shapes of the curves corresponding to different mass ratios remain mostly unchanged. In contrast, varying the chirp mass results in a shift along $\tau_\mathrm{ratio}/\tau_\mathrm{target}$. This shift scales as $\mathcal{M}^{-5/3}$, consistent with the dependence of the waveform duration on chirp mass.
We verify this behavior by repeating the analysis for chirp masses of $0.871\,M_\odot, 1.741\,M_\odot,$ and $2.612\,M_\odot$, as shown in Fig.\ref{fig:different_chirp_mass_error}. When an amplitude-based truncation threshold is applied, the resulting \snr accuracy is consistent across all tested chirp masses.

\bibliographystyle{apsrev4-1}
\bibliography{references}

\end{document}